# Elemental Substitution Effect and Electrochemical Reaction Dependent Surface Reconstruction in High-Entropy Alloy Electrocatalyst

Shreyasi Chattopadhyay[a], Raphael B. de Oliveira[b], Deepti Gangwar[c,d], Tymofii S Pieshkov[a], Marcelo L. Pereira Junior[a,e], Dhiman Banik[c,f], Astrid Campos-Mata[a], Atin Pramanik[a], Soumyabrata Roy[a], Douglas S. Galvão[g*], Chandra Sekhar Tiwary[f*], Krishanu Biswas[c*], Pulickel M Ajayan[a*]

[a]Department of Materials Science and NanoEngineering, Rice University, 6100 Main Street, Houston, Texas 77005, United States

[b]Department of Physics, Federal University of Rio Grande do Norte, Natal, RN, Brazil

[c]Department of Material Science and Engineering, Indian Institute of Technology, Kanpur, Uttar Pradesh, India 208016

[d]School of Engineering & Technology, Maharishi University of Information & Technology, Noida campus, Gautam Buddha Nagar, India

[e]College of Technology, University of Brasília, Brasília, DF, Brazil

[f]Department of Metallurgical and Materials Engineering, Indian Institute of Technology, Kharagpur, West Bengal, India 721302

[g]State University of Campinas, Campinas, SP, Brazil

**ABSTRACT:** High-entropy alloys (HEAs) offer a fascinating platform for electrocatalysis through their compositional complexity and tunable surface chemistry. Moreover, such structural complexity also leads to the reaction-specific surface reconstruction governing the stability and activity of the alloy system. The present study investigates the influence of elemental substitution on alkaline water electrolysis activity and the electrochemical-environment dependent surface restructuring in MnFeCoNiCu HEA under hydrogen (HER) and oxygen (OER) evolution reaction conditions. Cr-to-Cu substitution was done for MnFeCoNi-X system and activities were investigated toward alkaline HER and OER. Under identical conditions, MnFeCoNiCu (HEA-Cu) exhibits superior bifunctional activity than MnFeCoNiCr (HEA-Cr), requiring overpotentials of 538 and 460 mV at 10 mA $cm^{-2}$ for HER and OER, respectively. Furthermore, the restructuring behavior investigation reveals Cu migration and formation of a Cu-rich outer shell surrounding the multimetallic core after OER, resulting in activity degradation over time. Whereas such reconstruction is absent for HER, where surface activation and promotion of catalytic performance was observed. Computational analysis reveals that Cr-to-Cu substitution modulates the electronic structure and promotes more favorable adsorption of H*, O*, OH*, and OOH* intermediates. These findings establish a direct composition-structural chemistry-activity relationship and reaction-specific surface engineering of active-site in HEA electrocatalysts.

## INTRODUCTION

High-entropy alloys (HEAs), composed of multiple principal elements distributed within a single or near-single-phase solid solution, have emerged as a promising catalyst platform. The compositionally complex atomic environments of HEAs enable simultaneous modulation of electronic structure, adsorption energetics, lattice strain, charge transfer, and surface reconstruction.[1–3] Unlike conventional mono- or bimetallic catalysts, HEAs still remain an interesting yet mysterious alloy system. It provides a broad compositional design space in which neighboring atomic configurations can generate diverse catalytic sites and tune the adsorption of reaction intermediates. The catalytic activity of a HEA often arises from the interplay between the intrinsic activity of the individual elements and the synergistic effects generated at the interfaces within its structure. Understanding this interplay includes designing high-performance HEA via compositional optimization for favorable adsorption/desorption of reaction intermediates subject to Sabatier relations which is crucial for electrocatalysis.[4–7] Related studies on FeCoNiCrZn/Mn HEAs also showed that replacing one transition-metal component can shift the d-band characteristics and modify electrochemical behavior through multielement electronic synergy.[8] In addition, nano-structuring of multicomponent Fe-Co-Ni-Cu-Zn HEAs has been reported to increase accessible surface area and alter the electrochemical response, emphasizing that catalytic performance is governed not only by composition but also by particle size, defect density, surface oxidation, and elemental distribution.[9] HEA with a theoretical composition of $Ni_{14}Co_{14}Fe_{14}Mo_{6}Mn_{52}$ has been synthesized for HER and OER bifunctional catalytic activity for efficient water splitting.[10] Another theoretical calculation demonstrated that Ni exhibits a downshifted d-band center once alloyed by Cu, which promoted the desorption of H from the Ni surface to improve the catalytic activity for hydrogenation reactions.[11] In another work, four adjacent 3d-metal elements (Fe, Co, Ni, and Mn) were assembled to form an HEA without phase separation. The HEA with a high d-band center where Mn has the largest atomic radius among the composition was found to be a face-centered cubic (fcc) phase without phase separation and it showed substantial OER catalytic activity which was further enhanced after modification of d-band center using Pt and Ir.[12] In our previous work on replacing Mn by Zn in Fe-Co-Ni-Cu system,[13] compositional variation in transition-metal HEAs was shown to strongly influence bifunctional water-splitting activity. In another work on FeCoNiMoM (M=Cr, Cu) high-entropy material, composition dependent variation in alkaline OER activity was report.[14]

All these findings along highlight the fact that the single-element substitution or phase structural modification in the multicomponent alloy can substantially alter the redox accessibility and electrochemical kinetics of the active surface.[15] However, investigation of active metal site, interaction with reaction intermediates and surface restructuring under alkaline water electrocatalysis conditions are still limited. Among transition-metal HEAs, the CrMnFeCoNi alloy is one of the most extensively studied model systems because of its relatively stable face-centered cubic structure, compositional homogeneity that allows scope to investigate structure-property relationships.[8,16–19] In particular, multivalent Cr is generally associated with the formation of stable oxide/hydroxide surface layers and improved structural durability, whereas replacing monovalent Cu may promote electronic conductivity, local compositional heterogeneity, and modify adsorption/desorption kinetics of the redox intermediates. Therefore, comparing CrMnFeCoNi (HEA-Cr) and MnFeCoNiCu (HEA-Cu) provides an opportunity to establish composition-structure-electrocatalytic activity relationships in structurally related HEAs, yet to be reported. The current study focuses on the mechanistic role of single-element substitution in regulating structural chemistry and electrocatalytic performance, along with structural reconstruction under electrocatalysis conditions. Such a systematic study can clarify how Cr-to-Cu substitution affects phase constitution, electronic structure, electrochemically active surface area, reaction kinetics, surface reconstruction, and long-term stability, thereby guiding the rational design of low-cost multicomponent alloy catalysts for efficient water electrolysis.

## RESULTS AND DISCUSSION

### Structural analysis

The equiatomic FeCoNiCrMn alloy nanoparticles were synthesized via arc melting followed by cryo-milling. Room-temperature X-ray diffraction (XRD) analysis confirmed the formation of a single-phase face-centered cubic (FCC) solid solution in both the bulk alloy and the cryomilled powder, crystallizing in the Fm-3m space group. The diffraction pattern exhibited characteristic reflections at $2\theta \approx 43.5°$, 50.6°, 74.4°, 90.3°, and 95.6°, corresponding to the (111), (200), (220), (311), and (222) crystallographic planes, respectively (**Figure S1**). These diffraction peaks agree well with the standard reference pattern for the Co-Cr-Fe-Mn-Ni alloy system (JCPDS No. 04-029-1381). They are consistent with previously reported results for as-cast CoCrFeMnNi alloys.[20–22] **Figure 1**a presents the XRD pattern of the synthesized HEA-Cu nanoparticles. Subsequent substitution of Cr with Cu did not induce any phase transformation, and the alloy retained its single-phase FCC structure. Notably, the systematic shift of the diffraction peaks indicates that although Cu inclusion into the FeCoNiMn matrix did not change the crystal structure, it did cause a modest adjustment of the lattice parameters, resulting from the lattice distortion effect and which might also influence the high-entropy effect.[11] These observations confirm the successful formation of a Cu-containing FCC HEA while preserving the structural integrity of the parent alloy.

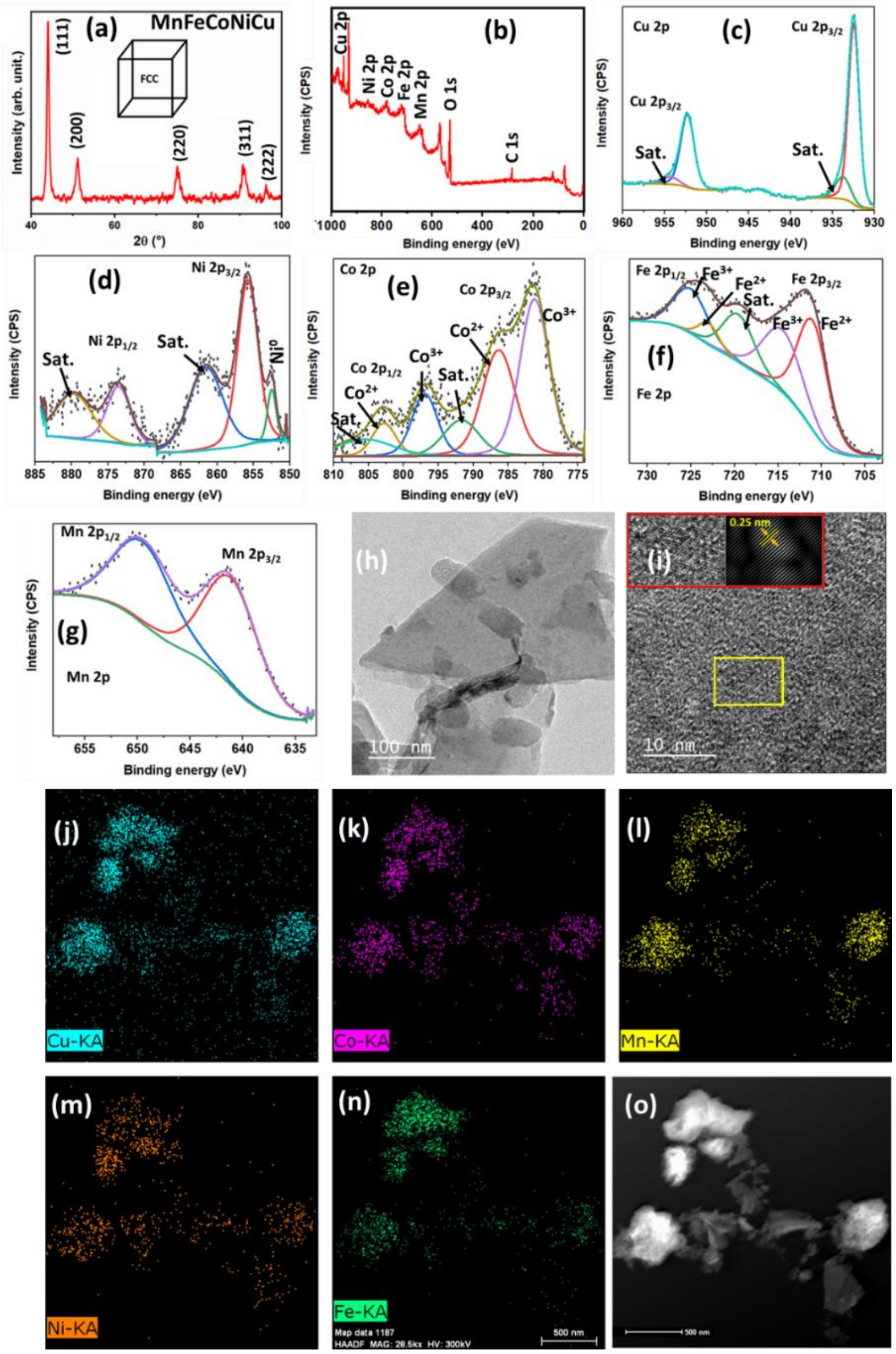

**Figure 1.** Characterization of MnFeCoNiCu: (a) XRD; (b-g) XPS; (h-i) TEM analysis (j-o) TEM elemental mapping.

The elemental composition and oxidation states of the HEA-Cu high-entropy alloy nanoparticles were further examined by X-ray photoelectron spectroscopy (XPS), as illustrated in **Figure 1**b-g. The survey-scan XPS spectrum (**Figure 1**b) confirms the presence of Co, Cu, Fe, Mn, and Ni. The high-resolution Cu 2p XPS spectrum (**Figure 1**c) exhibits two distinct peaks centered at 932.5 and 952.3 eV, corresponding to the Cu $2p_{3/2}$ and Cu $2p_{1/2}$ spin-orbit components, respectively.[23] The satellite peaks at 933.2 and 954.2 eV are associated to Cu 2p. The dominant peaks at 932.5 and 952.3 eV suggest that copper predominantly exists in the Cu(I) ($Cu^{+}$ oxidation state). Furthermore, the Cu LMM Auger peak observed at approximately 570 eV in the survey spectrum (Figure 1b) provides additional evidence for the presence of $Cu^{+}$ species. $Cu^{2+}$ is characterized by weak shake-up satellite features that appear in the 940-945 eV region. As shown in **Figure 1**d, the high-resolution Ni 2p spectrum can be deconvoluted into four components. The peaks located at 886.2 and 872.8 eV are assigned to the Ni $2p_{3/2}$ and Ni $2p_{1/2}$ levels, respectively, indicating the presence of $Ni^{2+}$ species. The peaks at 887.7 and 861.9 eV correspond to characteristic satellite features of Ni 2p. In addition, a peak observed at 852.2 eV is attributed to metallic $Ni^{0}$ (Ni $2p_{3/2}$), suggesting a minor fraction of elemental nickel within the alloy nanoparticles.[24] The Co 2p spectrum (**Figure 1**(e)) is resolved into four distinct peaks. $Co^{3+}$ species have peaks at 780.4 eV (Co $2p_{3/2}$) and 796.7 eV (Co $2p_{1/2}$), while $Co^{2+}$ has peaks at 786.1 and 803.4 eV (Co $2p_{3/2}$ and Co $2p_{1/2}$, respectively).[25] Two satellite peaks are also present. These results indicate the coexistence of both +2 and +3 oxidation states of cobalt on the nanoparticle surface. Fe 2p XPS (**Figure 1**f) spectrum shows two prominent peaks for Fe $2p_{3/2}$ and Fe $2p_{1/2}$, which are deconvoluted into four peaks and a satellite peak, indicating the coexistence of Fe in +3 and +2 valence states.[26] No peak for Fe metal is found. The high-resolution Mn 2p spectrum shown in **Figure 1**g exhibits two well-defined peaks centered at 645.8 and 659.4 eV, corresponding to the Mn $2p_{3/2}$ and Mn $2p_{1/2}$ spin–orbit components, respectively.[27] These binding energies are characteristic of $Mn^{4+}$ species, indicating that manganese predominantly exists in the +4 oxidation state within the alloy nanoparticles. Overall, the XPS analysis reveals that Cu, Ni, and Mn are present predominantly in single oxidation states, namely $Cu^{+}$, $Ni^{2+}$, and $Mn^{4+}$, respectively. In contrast, Fe and Co exhibit mixed-valence characteristics, suggesting the coexistence of multiple oxidation states on the nanoparticle surface. Detailed XPS analysis of the HEA-Cr alloy shown in **Figure S2** demonstrates predominant presence of $Ni^{2+}$, $Co^{2+/3+}$, $Mn^{4+}$, $Fe^{2+/3+}$ and $Cr^{3+/6+}$. Overall, the XPS analysis reveals that Ni, and Mn are present predominantly in single oxidation states, namely, $Ni^{2+}$, and $Mn^{4+}$, respectively. In contrast, Fe, Co and Cu exhibit mixed-valence characteristics, suggesting the coexistence of multiple oxidation states on the nanoparticle surface.

The microstructural characteristics of the HEA-Cr and HEA-Cu high-entropy alloys were further investigated. SEM analysis of the HEA-Cu (**Figure S3**) and HEA-Cr (**Figure S4**) shows homogeneous distribution of the elements. Furthermore, the low- and high-magnification HRTEM images are presented in **Figure 1**(h–i) and **Figure S5**(a). The high-resolution image of HEA-Cu in **Figure 1**i reveals well-defined lattice fringes, indicative of the high crystallinity of the synthesized nanoparticles. The measured interplanar spacing of approximately 0.25 nm, corresponds to the (111) crystallographic plane of the FCC HEA structure. The absence of additional lattice features or secondary phases confirms the formation of a single-phase alloy, in agreement with the XRD results. Furthermore, the high-angle annular dark-field scanning transmission electron microscopy (HAADF-STEM) image (**Figure 1**o) and the corresponding elemental mapping images (**Figure 1**j–n) demonstrate a uniform spatial distribution of Co, Cu, Fe, Mn, and Ni throughout the nanoparticles. No evidence of elemental segregation or agglomeration is observed, confirming the homogeneous incorporation of all constituent elements within the HEA matrix.

### Electrocatalytic Performances and Reaction dependent Restructuring

Activity of HEA-Cr and HEA-Cu was studied for both HER and OER in 1M alkaline water electrolysis. To understand the actual process, especially structural evolution involves in multi metallic HEA system under experimental conditions of electrocatalysis, set of testing was done using the catalyst ink prepared without adding any conducting carbon. Another set of electrocatalytic study was also performed using catalyst ink prepared with 10% Vulcan carbon for activity comparison. Optimization of ink loading was also done and the best result for homogeneous catalyst loading on GCE surface was found to be 0.5mg $cm^{-2}$.

Linear sweep voltammetry (LSV) polarization curves for HER, presented **Figure 2**a showed metallic Cu required an overpotential of 786 mV vs RHE to achieve a current density of 10 $mAcm^{-2}$. In contrast, the onset potential for CrMnFeCoNi (HEA-Cr) and MnFeCoNiCu (HEA-Cu) were observed to be 601 and 538 mV vs RHE, demonstrating superior performance for high entropy alloy system than single component metal, and further improvement via replacing Cr by Cu. Tafel analysis is a valuable tool for elucidating electrocatalytic kinetics and identifying the potential rate-determining step (RDS) of the reaction.[6,28] The reaction mechanism for both materials was further studied through Tafel analysis, as shown in **Figure 2**b. The Tafel slopes obtained were 226 mVdec$^{-1}$, 184 and 165mVdec$^{-1}$ for Cu, HEA-Cr and HEA-Cu, respectively. These values indicate that the Volmer-Heyrovsky mechanism is the rate determining step in the reaction. These results further demonstrate the advantage of the synergy of the five metals in the high entropy alloy improving the catalytic HER activity compared to single Cu metal catalyst. Moreover, replacing Cr with Cu in HEA structure facilitates the kinetics and result in enhancement in activity. To assess the structural stability and surface chemistry of HEA-Cu, accelerated durability tests (ADT) were conducted. After 3000 cycles under applied potential for HER, an increase in the catalytic activity was observed, as presented in **Figure 2**c-d. The onset potential decreased by 22.8 mV.

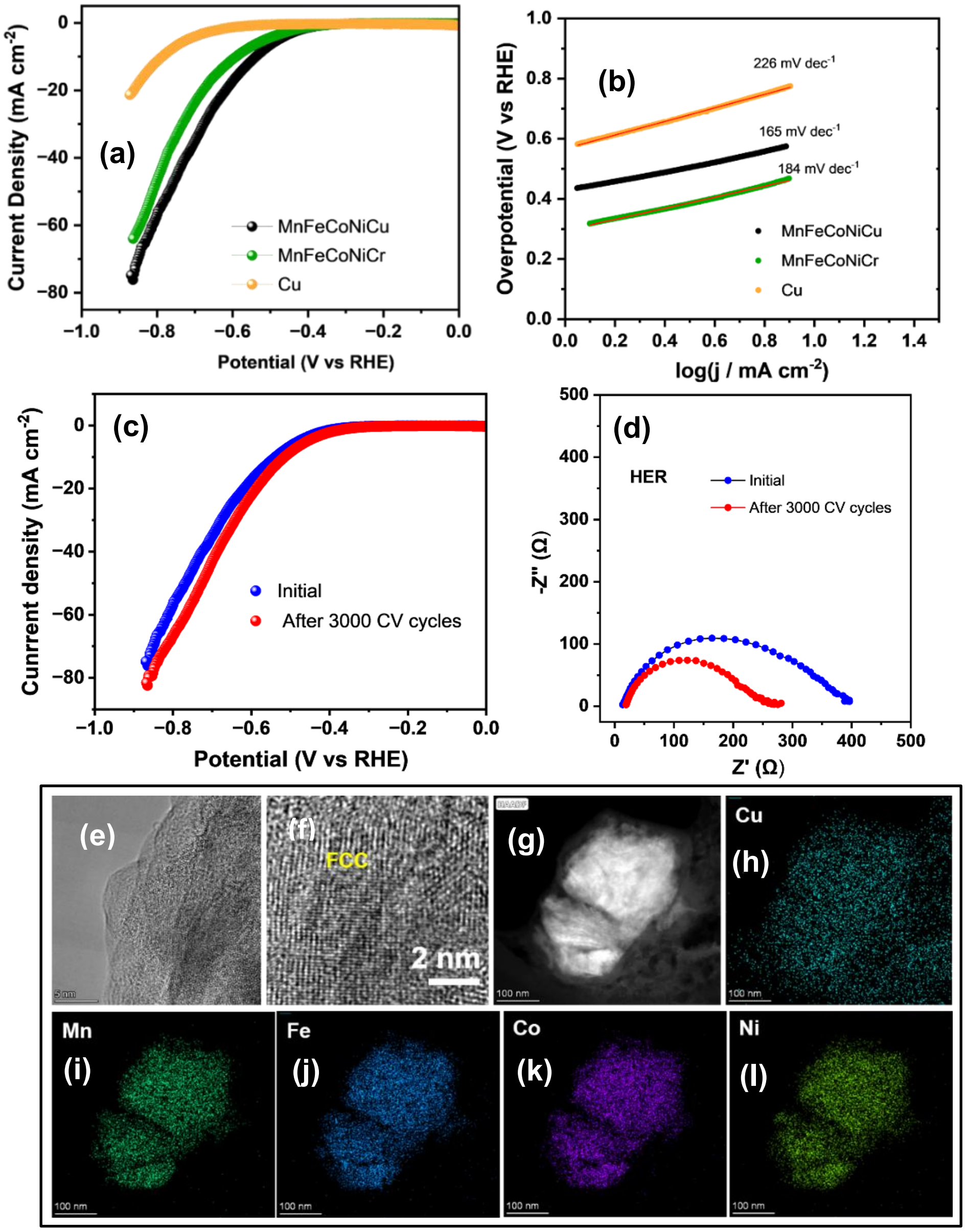

**Figure 2.** HER activity of HEAs in 1M KOH: (a, b) LSV curves and Tafel plot of HEAs and Cu, respectively. (c, d) LSVs for MnFeCoNiCu (HEA-Cu) before and after ADT, and corresponding EIS, respectively. (e, f) TEM analysis of the recovered HEA-Cu catalysts after ADT. (g-l) TEM-EDS mapping of the recovered HEA-Cu catalysts after ADT.

Electrochemical Impedance Spectroscopy was carried out to determine the charge-transfer resistance ($R_{CT}$) values for the HEA, both before and after 3000 cycles. The $R_{CT}$ values (Figure 2d) obtained were 339 Ω and 208 Ω before and after 3000 cycles, respectively, suggesting activity enhancement by decrease in charge transfer resistance. This improvement can be attributed to the structural stabilization of the HEA and possible activation of the surface. TEM analysis of the recovered catalysts after HER was further done to investigate the structural change/reconstruction that might influence HER activity behavior.

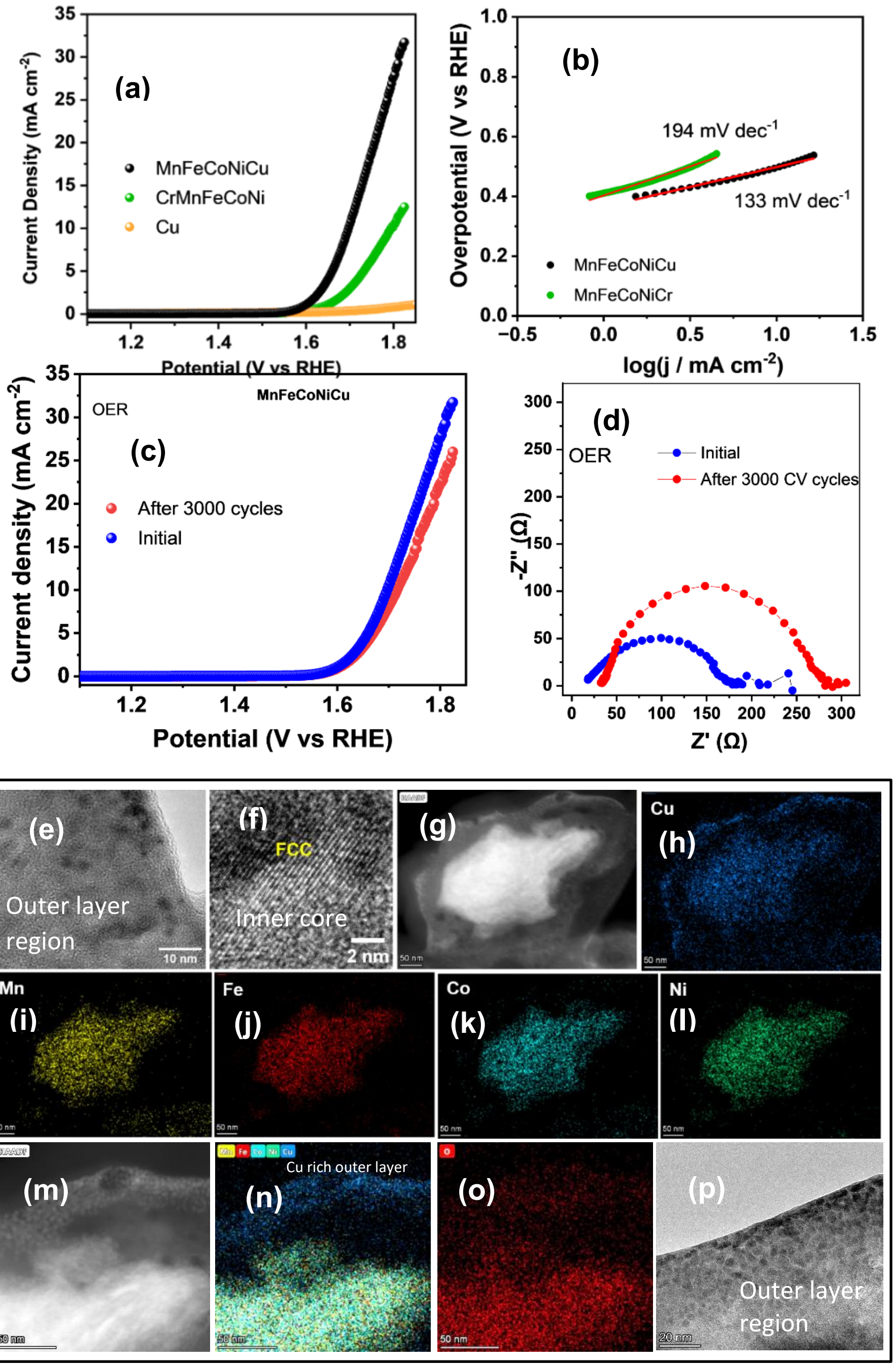

**Figure 3.** OER activity of HEAs in 1M KOH: (a, b) LSV curves and Tafel plot of HEAs and Cu, respectively. (c, d) LSVs for MnFeCoNiCu (HEA-Cu) before and after ADT, and corresponding EIS, respectively. (e, f) TEM analysis of the recovered HEA-Cu catalysts after ADT. (g-o) TEM-EDS mapping of the recovered HEA-Cu catalysts after ADT. (p) Magnified portion showing the outer layer.

The corresponding TEM results of the recovered HEA-Cu after HER are shown in **Figure 2**e-l and **S5**a. It is noteworthy that crystalline structure of the HEA catalyst retained after the ADT testing as can be seen from the fringes present in **Figure 2**f. EDS mapping on recovered catalyst after HER showed homogeneous distribution of the existing elements (Figure 2g-l) without significant separation proving the structural stability. The enhanced activity could be attributed to surface activation arising from changes in the local chemical environment under reductive HER conditions, driven by the applied potential and interactions with HER intermediates.

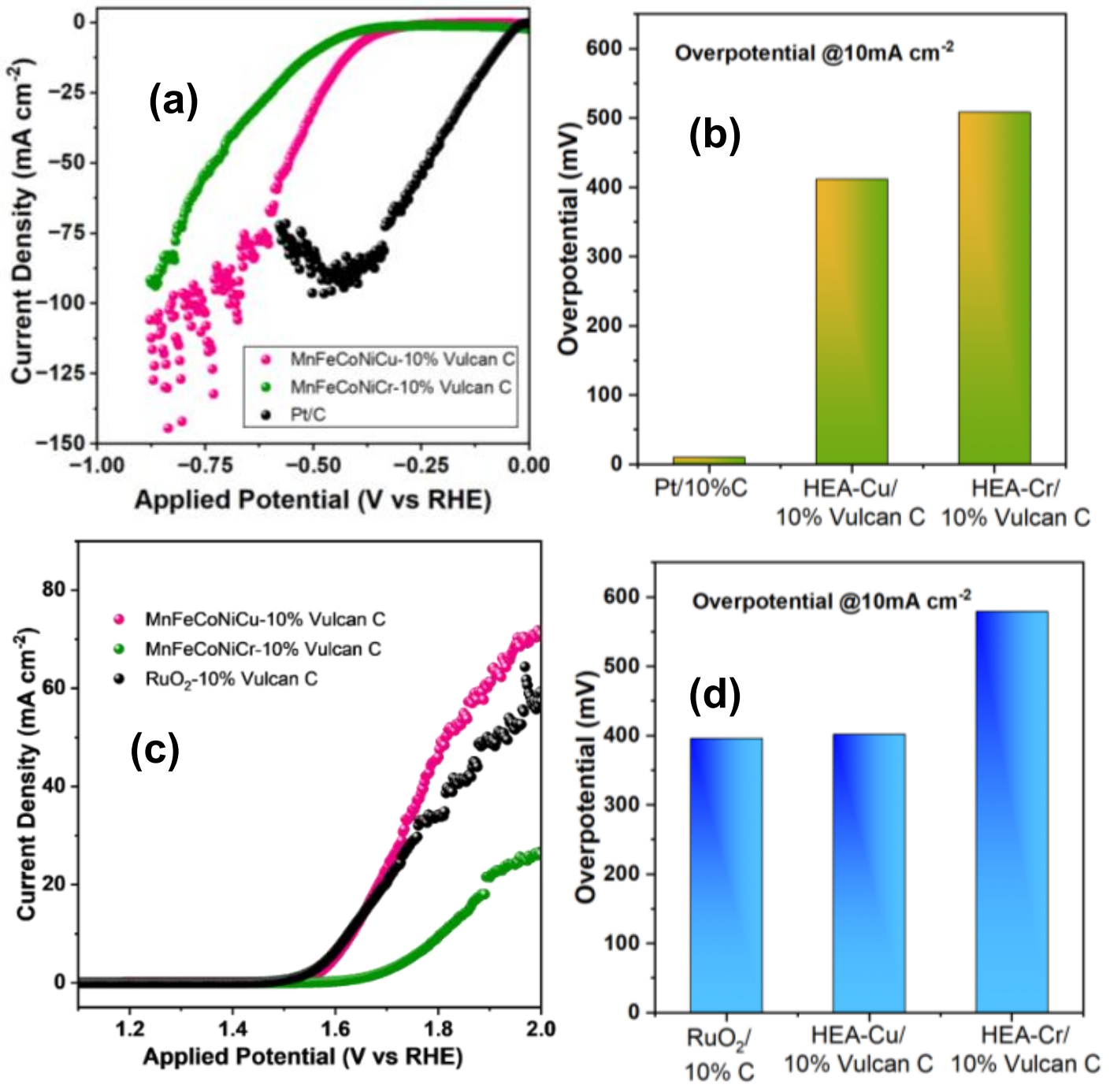

**Figure 4.** HER and OER activity of HEA-Cu, HEA-Cr ink prepared with mixing 10% Vulcan C in 1M KOH: (a, b) LSV curves and overpotential of HEAs, respectively for HER; (c, d) LSV and overpotential of HEAs OER, respectively. Comparisons with $RuO_2$ for OER and Pt/C for HER are also included.

Electrocatalysis activity testing results under OER reaction condition are shown in **Figure 3**a, b. The overpotential values at 10mA cm$^{-2}$ were found to be 569 and 460 mV vs RHE for HEA-Cr and HEA-Cu, respectively. Almost no activity was seen for Cu as catalyst. This result, HEA-Cu showing the lowest overpotential at 10mA cm$^{-2}$ and lower Tafel slope than HEA-Cr, suggest scope for improvement in activity via compositional modification. Electrochemical Active Surface area (ECSA) value of HEA-Cu was further evaluated by determining the double layer capacitance ($C_{dl}$) value (**Figure S6**). Corresponding $C_{dl}$ and ECSA values were found to be 0.12 mF cm$^{-2}$ and 3 cm$^2$. To assess the structural stability and surface chemistry of HEA-Cu, ADT were for 3000 cycles under applied potential for OER. A decrease in current density value was observed after 3000 cycles at higher potential region (**Figure 3**c). However, no significant change in overpotential @10 mA cm$^{-2}$ was observed. **Figure 3**d shows the Nyquist plots for OER, an increase in $R_{CT}$ values before and after ADT in EIS analysis (Figure 3d) corroborates with the finding obtained from LSV. TEM analysis of the recovered catalysts after OER were further done to investigate the reason behind such activity change, and any nano-structural change/reconstruction that might influence this OER activity. TEM analysis results of the recovered HEA-Cu after ADT under OER condition are shown in **Figure 3**e-p and **S5**b where formation of an outer layer with a hint of particle formation can be seen. It is noteworthy that the outer layers is amorphous and the crystalline structure of the HEA catalyst retained in the core after the ADT testing as can be seen from the fringes present in **Figure 3**f. Interestingly, unlike HER, formation of Cu rich outer layer **(Figure 3**g-l) was found for the catalyst recovered after OER, suggesting restructuring under testing condition. Notably, Cu found to exist on the outer layer (**Figure 3**h, n) and the central part was found to contain all five metals including Cu. These results suggest that under OER conditions, Cu tends to migrate to the surface resulting in surface structural reconstruction. Moreover, Cu tends to form oxide easily and that could affect the interaction of the OER intermediates with the surface, especially such surface can bind the oxygen containing intermediates more strongly leading to slower kinetics and decrease in activity. That could be the probable reason for the decrease in activity after 3000 cycles.

To demonstrate practical applicability and enable comparison with literature reports in which Vulcan carbon is usually incorporated into catalyst inks, the HER and OER activities of the HEA catalysts were also evaluated. A better kinetics were observed with increase in overall current density values with minimal decrease in overpotential at 10 mA $cm^{-2}$, while testing HEA-Cr and HEA-Cu in 1 M KOH using inks containing 10 wt% added Vulcan carbon (**Figure 4**a-d). The overall trend was found to be same as observed from the results obtained from no carbon added catalyst inks and other experimental conditions, HEA-Cu is better than HEA-Cr. Moreover, the data noise present in the LSVs are from the bubble formation, which also direct evidence of superior activity of HEA-Cu. Notably, addition of Vulcan carbon improves HER and OER performance by improving conductivity, charge transfer, catalyst dispersion, and active-site utilization. While comparing the HER overpotential values, Pt/C catalyst outperformed the HEA catalysts under same experimental condition. However, for OER, HEA-Cu showed similar activity to $RuO_2$, even higher activity at higher overpotential, demonstrating the potential to replace the noble metal catalyst.

**Probable Origin of Restructuring**

The restructuring trend observed is schematically represented in **Scheme 1**. HER condition did not allow any metal-specific migration therefore showed retention of homogeneous uniform distribution (**Scheme 1**a) of the elements. The change in elemental distribution during OER (**Scheme 1**b) can be due to high affinity of Cu with oxygen which results in diffusion of Cu atoms in oxygen rich intermediate environment. For OER, we observe Cu and oxygen rich surface (**Scheme 3**n, o) followed by a Cu depleted region which suggests diffusion control reaction. The preferential accumulation of Cu at the outer surface of HEA-Cu following OER can be rationalized by the chemistry between elemental surface segregation, electrochemically induced oxidation, and potential-dependent atomic redistribution. Despite its comparatively lower oxygen affinity, Cu possesses a relatively low metallic surface energy, which may favor its segregation toward the alloy surface. Under anodic polarization, interactions with hydroxyl and oxygen-containing species can further facilitate Cu oxidation and atomic mobility, promoting the development of a Cu-rich surface layer. Simultaneous selective oxidation, dissolution, or redistribution of other constituent metals may additionally contribute to the observed surface enrichment. Notably, a study on Cr containing Cantor alloy reported initial formation Cr-oxide surface layer followed by exfoliation leading to total activity loss.[29] However, our Cu system showed different tendency and no total loss of activity was observed. This also suggests the composition dependent complex behavior of restructuring tendency of HEAs. In case of our HER observation supported by TEM-EDS mapping, such layer does not form in the reducing environment of HER due to low affinity of hydrogen with Cu. The cathodic environment during HER favors reduced surface states and does not provide the same oxidation-assisted driving force for Cu-rich shell formation. Consequently, the multimetallic distribution remains comparatively homogeneous after HER, accompanied by surface activation and improved catalytic performance. These findings reveal reaction-environment-dependent surface evolution in MnFeCoNiCu HEA, where oxidative reconstruction correlates with OER degradation, whereas reductive conditions preserve the multimetallic structure and promote HER activity.

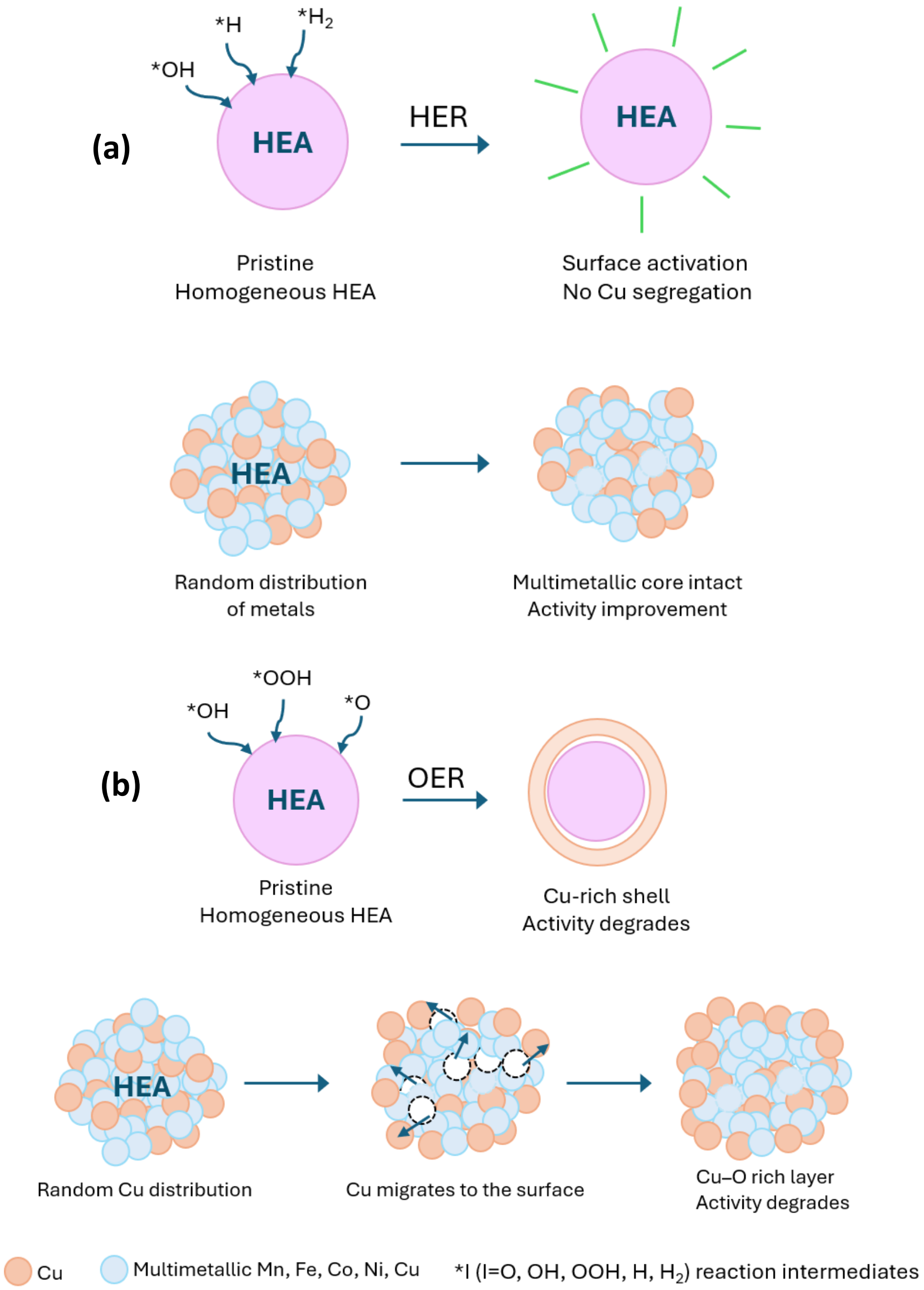

**Scheme 1.** Schematic representation showing the surface restructuring during (a) HER and (b) OER.

## Computational Analysis

Detail computational investigation was conducted to support our experimental observations and determine the structure-property relationship. Our analysis began with optimizing all structures. As we had a large number of structures to optimize, it was necessary to adopt certain quality parameters for which structures we would use for catalytic activity. Thus, at the end of the optimization of the two-hundred structures (100 for HEA with Cu and 100 for HEA with Cr), we organized them by increasing energy stability. During structural optimization, we also allowed the lattice parameters to change. In general, the HEA-Cr systems preserved a more cohesive FCC framework, with the lattice angles α, β, and γ deviating by approximately 0.1% from the ideal cubic value of 90°. In contrast, the HEA-Cu HEA structures exhibited pronounced distortions, with at least one lattice angle varying by 10-15%. This behavior is consistent with the difference in the van der Waals radii of Cu and Cr, since Cr presents a van der Waals radius approximately 10% larger than that of Cu, which can influence local atomic packing and the resulting structural relaxation.[30] This allows for a more cohesive structure during optimization.

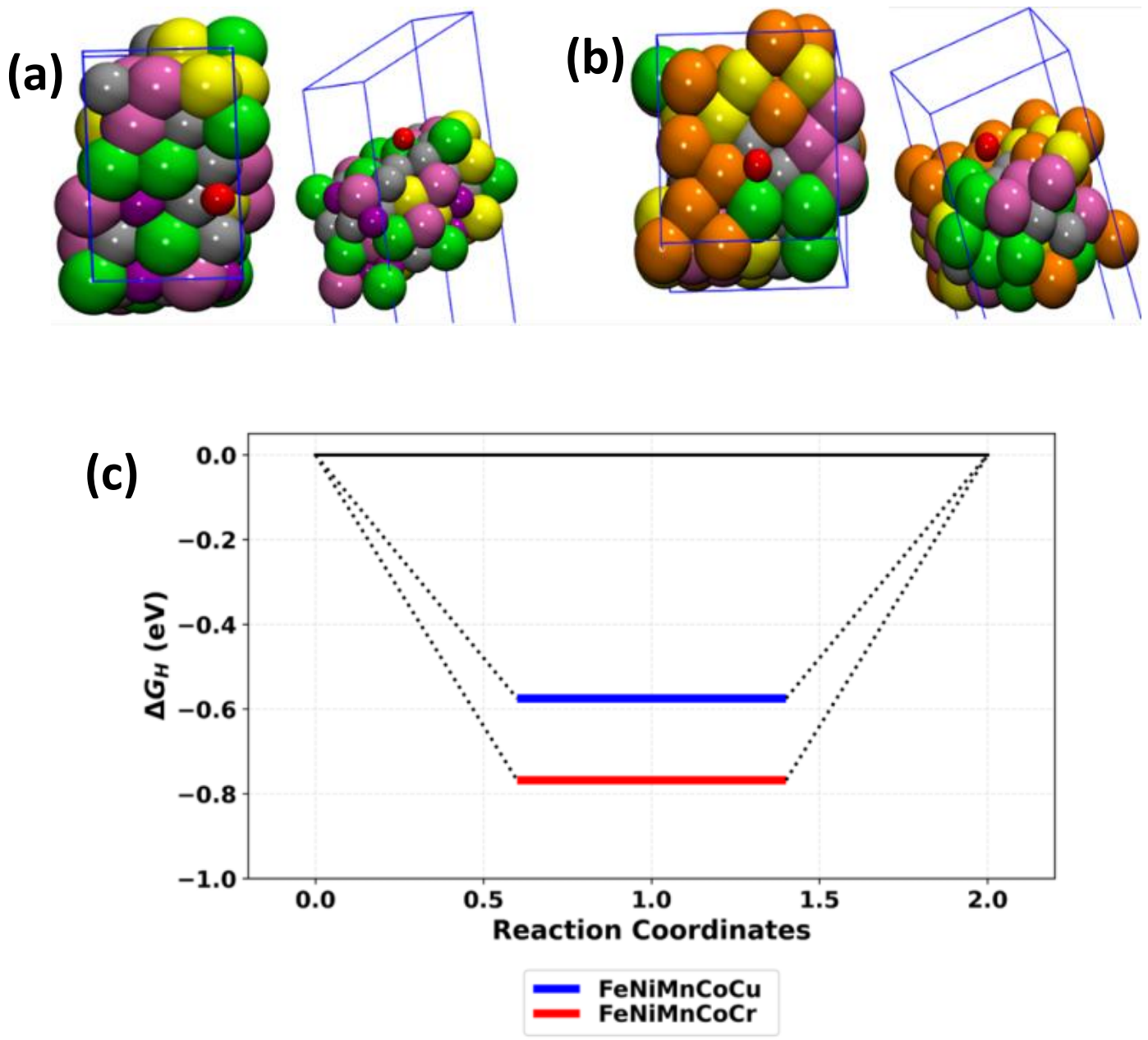

**Figure 5.** (a) FeNiMnCoCu and (b) to FeNiMnCoCr, with the H atom placed at the most favorable adsorption site; (c) Gibbs Free Energy for each one of the structures. Atomic species are represented as Fe (yellow), H (red), Cu (purple), Mn (green), Co (mauve), Ni (silver), and Cr (orange).

This can result in less atomic movement during the optimization process. We focus this analysis on the lattice angles, since the lattice vectors depend more on the initial atom distribution. Since all structures were previously ranked according to their energetic stability, the five most stable configurations of each HEA type, HEA-Cu and HEA-Cr, were selected to evaluate their HER performance. Adsorption sites for H were identified using a Monte Carlo sampling approach, allowing the determination of the lowest-energy adsorption configuration for each of the ten selected structures. After placing H at the most favorable adsorption site (**Figure 5**a, b), the hydrogen adsorption free energy ($\Delta G_H$, see Eq. 5) was estimated considering T = 300 K, pH = 14, and U = −0.80 V. Figure 6 and Figure SI present the ten structures with the H atom positioned at the energetically preferred adsorption site obtained from the Monte Carlo calculations. In Figure 6, we display the two structures with the best value of the $\Delta G_H$; meanwhile, in Figure SI, we display the rest of the configurations of HEA-Cu and HEA-Cr HEA. In case of the HEA-Cr HEA, H preferentially remains in Cr-rich regions of the surface and, in some configurations, is located nearly on top of a Cr atom, consistent with the electron-donor character commonly associated with Cr. This behavior has been related to facilitated water dissociation and multisite synergistic effects in Cr-containing HEAs.[31] In contrast, for the HEA-Cu HEA, even when the most favorable adsorption site lies within a Cu-rich region, the H atom tends to remain relatively distant from the nearest Cu atom, typically occupying bridge or hollow positions adjacent to Ni, Co, or Fe atoms, thus avoiding direct top-Cu binding. This trend is consistent with the higher electronegativity of Cu compared to other 3d transition metals, leading Cu to behave as an electron-acceptor species that mainly modulates the electronic environment of neighboring atoms rather than acting as the preferred adsorption center for H.[7,32] **Figure 5**c shows a comparison between the $\Delta G_H$ value for the HEA-Cu and HEA-Cr third structures, and Table SI presents the rest of those values. **Figure 5**c reveals that $\Delta G_H$ is less negative for HEA-Cu, indicating weaker hydrogen binding than for HEA-Cr. Notably, the third-most-structurally-stable configuration in each HEA set has the highest $\Delta G_H$ value, resulting in a difference of approximately 0.2 eV between the two systems. According to the Sabatier principle,[33,34]the reaction intermediate (H* in this case) should exhibit moderate binding strength to favor efficient desorption and enhanced HER kinetics.[7]

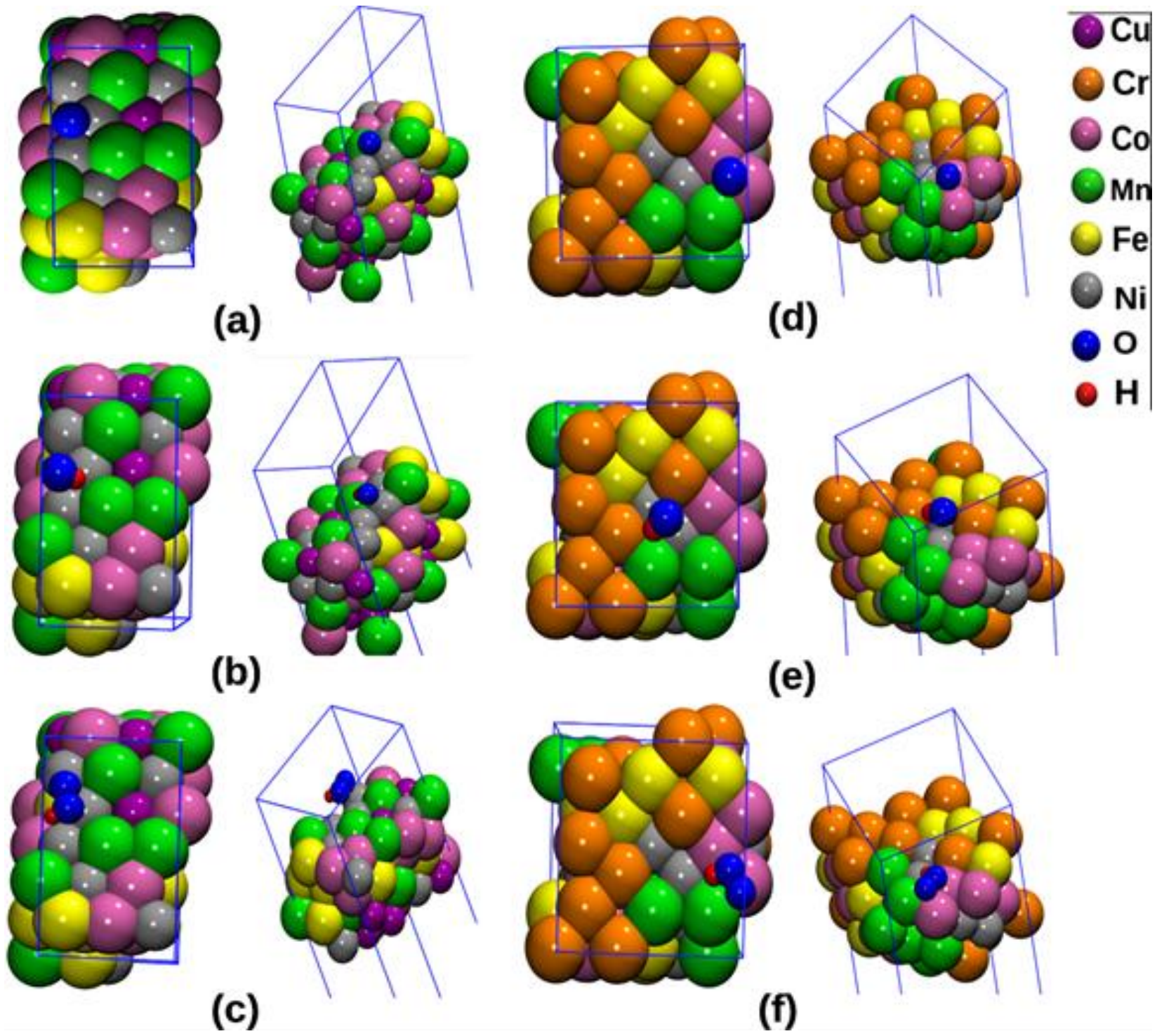

**Figure 6.** FeNiMnCoCu (a), (b), and (c), and FeNiMnCoCr (d), (e), and (f) structures selected from the OER analysis, with the O, OH, and OOH intermediate, respectively, placed at the most favorable adsorption site. Atomic species are represented as Fe (yellow), H (red), Cu (purple), Mn (green), Co (mauve), Ni (silver), Cr (orange), and O (blue).

The obtained results are consistent with the experimental trend, in which HEA-Cu shows higher catalytic activity than HEA-Cr. Cu-containing HEAs have been frequently reported as highly active HER catalysts, where Cu contributes to the formation of optimally bound H* intermediates through electron-acceptor behavior and d-band modulation effects.[7,32,35]In contrast, Cr plays a different role: although its electron-donor character may facilitate water dissociation through cooperative multi-site interactions, Cr-rich systems tend to form $Cr_2O_3$ passivation layers under alkaline conditions, which can reduce the intrinsic HER activity.[31]

For the OER calculations, one structure was selected for each HEA type, based on the configurations previously identified in the HER analysis. A Monte Carlo sampling approach was employed to identify the most favorable adsorption site for each intermediate molecule. **Figure 6**(a), (b), and (c) show, respectively, O, OH, and OOH in the HEA-Cu system, whereas **Figure 6**(d), (e), and (f) correspond to O, OH, and OOH in FeNiMnCoCr. In both HEAs, the preferred adsorption site for OOH differs from that observed for H adsorption. Nevertheless, a similar tendency is observed: in HEA-Cu, the intermediate molecules tend to be adsorbed on the energetically favored sites. In contrast, in HEA-Cr, the intermediate molecules remain adsorbed on Cr-rich surface regions, typically near Cr atoms. After full structural optimization, the Gibbs free energies of the intermediate molecules were estimated as $\Delta G_O$, $\Delta G_{OH}$, and $\Delta G_{OOH}$; the values are presented in Table 1. These results indicate that HEA-Cr binds more strongly than HEA-Cu with all the molecules considered during the OER process, consistent with the trend previously observed for H*. The stronger adsorption may influence the catalytic response by stabilizing intermediates more strongly, whereas the weaker binding observed for HEA-Cu is consistent with improved OER performance. This behavior aligns with experimental reports of higher OER activity in Cu-containing HEAs.

**Table 1.** ΔG (in eV) for the sub-reactions of the OER to the HEA-Cu and HEA-Cr structures.

| Molecule | $\Delta G_o$ | $\Delta G_{OH}$ | $\Delta G_{OOH}$ |
|---|---|---|---|
| **FeNiMnCoCu** | -1.0 | -1.03 | -2.27 |
| **FeNiMnCoCr** | -1.3 | -1.48 | -3.48 |

## CONCLUSION

The current work investigates the structural and electrochemical impacts of substituting Cr with Cu in MnFeCoNi high-entropy alloy (HEA) towards alkaline water electrolysis. Electrochemical evaluations in 1M KOH revealed that the Cu-substituted alloy (HEA-Cu) significantly outperformed its Cr-containing counterpart in both the HER and OER. Specifically, MnFeCoNiCu shows a lower HER onset potential of 538 mV at 10 mA $cm^{-2}$ and a Tafel slope of 165 $mVdec^{-1}$, and when tested with Vulcan carbon, its OER activity proved comparable to the noble-metal standard $RuO_2$. Computational and stability analyses further elucidate why the Cu-substituted HEA exhibits superior performance and how it behaves during prolonged electrolysis. Accelerated durability tests (ADT) showed that while HER conditions preserved a homogeneous elemental distribution, OER conditions triggered Cu migration to the surface, forming an oxygen-rich outer layer that can eventually lead to activity deterioration. DFT calculations supported these findings by showing that Cu substitution induces significant lattice distortions (10-15%) and acts as an electron acceptor to modulate the electronic environment of neighboring atoms. This electronic shift results in weaker, more optimal binding energies for reaction intermediates like H*, O*, OH*, and OOH*, which, facilitates more efficient adsorption and desorption kinetics for enhanced water splitting. The work clearly highlights the efficacy of HEA with controlled composition as building blocks for developing low cost, noble metal free sustainable catalyst for water splitting.

## Supporting Information

Synthesis and characterization, Computational methodology details, XRD of HEA-Cr (Figure S1), XPS of HEA-Cr (Figure S2), SEM analysis of HEA-Cu (Figure S3) and HEA-Cr (Figure S4), Low magnification TEM image of HEA-Cu HEA catalyst after HER and OER stability test (Figure S5), ECSA measurement plot (Figure S6).

**Table of Contents**

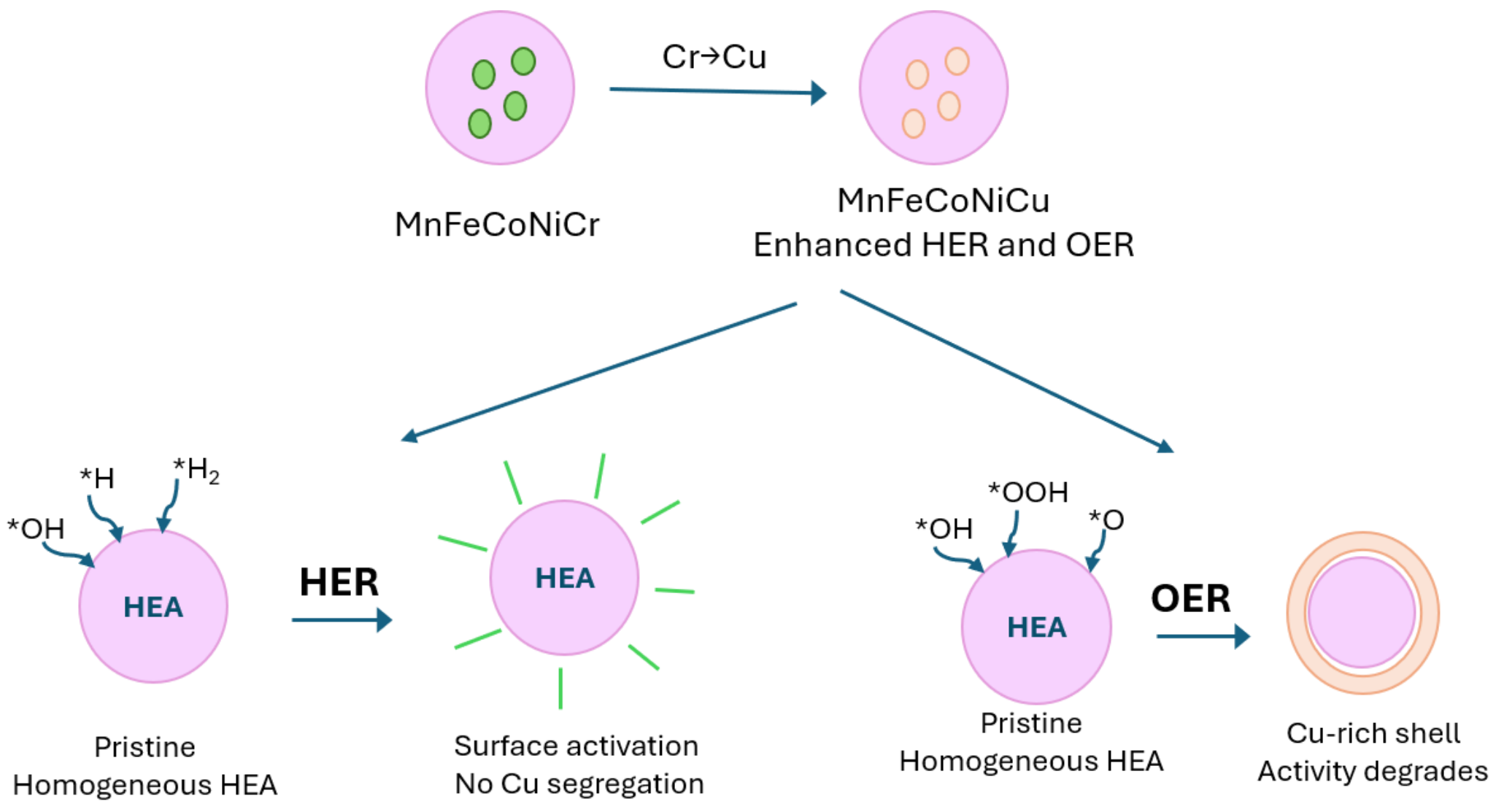

This study investigates elemental substitution and reaction-dependent surface reconstruction in MnFeCoNiCu high-entropy alloys. Replacing Cr with Cu enhances HER and OER activity. Post-OER characterization shows Cu migration forming a Cu-O rich shell, causing activity degradation, whereas HER promotes surface activation and improved catalytic performance without comparable reconstruction.

# Supporting Information

# Elemental Substitution Effect and Electrochemical Reaction Dependent Surface Reconstruction in High-Entropy Alloy Electrocatalyst

Shreyasi Chattopadhyay[a], Raphael B. de Oliveira[b], Deepti Gangwar[c,d], Tymofii S Pieshkov[a], Marcelo L. Pereira Junior[a,e], Dhiman Banik[c,f], Astrid Campos-Mata[a], Atin Pramanik[a], Soumyabrata Roy[a], Douglas S. Galvão[g*], Chandra Sekhar Tiwary[f*], Krishanu Biswas[c*], Pulickel M Ajayan[a*]

[a]Department of Materials Science and NanoEngineering, Rice University, 6100 Main Street, Houston, Texas 77005, United States

[b]Department of Physics, Federal University of Rio Grande do Norte, Natal, RN, Brazil

[c]Department of Material Science and Engineering, Indian Institute of Technology, Kanpur, Uttar Pradesh, India 208016

[d]School of Engineering & Technology, Maharishi University of Information & Technology, Noida campus, Gautam Buddha Nagar, India

[e]College of Technology, University of Brasília, Brasília, DF, Brazil

[f]Department of Metallurgical and Materials Engineering, Indian Institute of Technology, Kharagpur, West Bengal, India 721302

[g]State University of Campinas, Campinas, SP, Brazil

## 2. Experimental Section

### *2.1. Synthesis and Characterization*

High-entropy alloy (HEA) nanoparticles were synthesized using a melting followed by cryomilling process that enabled precise compositional control. The FCC homogeneous phase of the synthesized HEA has been confirmed using XRD. XPS data were collected on using Nexsa G2 XPS instrument and the spectra are calibrated with respect to C1s. Scanning electron microscopy (SEM) and Transmission electron microscopy (TEM) were performed for structural and compositional characterization using a JEOL IT800 and FEI Titan Themis3 microscope, respectively.

### *2.2. Electrochemical measurements*

Alkaline water electrolysis activity of the HEA was investigated with 1M KOH electrolyte and in a three-electrode set-up using Autolab Potentiostat. Catalyst ink loaded glassy carbon electrode (GCE) (surface area=0.07 $cm^2$, CH Instruments) was Pt and Hg/HgO (1M NaOH) were used as working, counter and reference electrode, respectively. To prepare the catalyst ink, 2 mg catalyst in was sonicated in 960 µL 2-Propanol and sonicated for 30 mins. Then 40 µL Nafion was added and sonicated for another hour. Finally, 20 µL ink was drop casted on GCE and dried at ambient condition. A control catalyst ink was also prepared with HEA before acid wash. An Autolab Potentiostat was used for cyclic voltammetry (CV), linear sweep voltammetry (LSV) and electrochemical impedance spectroscopy (EIS) measurement. LSV were recorded with a scan rate of 5 mV $s^{-1}$. EIS data were acquired in frequency range 100 kHz to 100 MHz with an amplitude of 10 mV. All the measurements were done in an ambient atmosphere, for HER experiment the KOH solution was purged Ar gas for 30min. The potentials were iR corrected and converted to reversible hydrogen electrode (RHE) according to Nernst equation: $E_{RHE}=E_{Hg/HgO}+E°_{Hg/HgO}+0.059pH$, ($E°_{Hg/HgO}=0.098V$). Calculation of electrochemical active surface area (ECSA) was done by determining the double-layer capacitance.

Electrocatalysis experiments were also conducted with the inks prepared with mixing 10% Vulcan carbon and same solvent mixture system.

### *2.3. Computational Analysis*

First-principles calculations were carried out within the framework of Density Functional Theory (DFT), detail is discussed in supporting information.

## Computational methodology

To comprehensively investigate the HER activity in the HEA, we employed the Computational Hydrogen Electrode (CHE) model.[1] In this model, we can analyze the HER mechanism under alkaline conditions, decomposing it into the following sub-reactions[2]:

$H_2O + e^- + * \rightarrow H^* + OH^-$ (Volmer), (1)

$H_2O + H^* + e^- \rightarrow H_2 + OH^-$ (Heyrovsky), (2)

and

$2H^* \rightarrow H_2$ (Tafel). (3)

In this notation, * denotes the active site on the HEA surface, and H represents a hydrogen atom adsorbed at this site. Regardless of the specific HER pathway, the Gibbs free energy change (ΔG) for hydrogen adsorption is widely used as a descriptor of catalytic activity. Within the Computational Hydrogen Electrode (CHE) framework, the Gibbs free energy of hydrogen adsorption is calculated as:

$$\Delta GH = EH - (E + (½)E_{H2}) + \Delta E_{ZPE} - T\Delta S, \quad (4)$$

where EH is the total energy of the HEA with an adsorbed H atom, E is the total energy of the clean HEA surface, and $E_{H2}$ represents the energy of an isolated $H_2$ molecule. $\Delta E_{ZPE}$ and ΔS correspond to the zero-point energy difference and the entropy change upon hydrogen adsorption relative to the gas phase, respectively. The effect of an applied potential (U) and the pH can be included by adding the terms −neU and kBTpH ln(10), where n is the number of transferred electrons (n = 1 for HER), e is the elementary charge ($1.602 \times 10^{-19}$ C), kB is the Boltzmann constant ($8.62 \times 10^{-5}$ eV/K), and T is the temperature. Rewriting the Gibbs free energy expression[3]:

$$\Delta GH = EH - (E + (½)E_{H2}) + \Delta EZPE - T\Delta S - neU + kBTpH\ln(10). \quad (5)$$

In this work, the $\Delta E_{ZPE}$ and TΔS contributions are approximated as 0.24 eV, following previous studies.[4]

We also seek to understand certain aspects of the oxygen evolution reaction (OER) mechanism through DFT calculations. Ideally, in an OER, we have a process of transferring 4 $e^-$, through 4 sub-reactions illustrated below[1]:

$$H_2O \rightarrow {}^*OH + H^+ + e^-, \quad (6)$$

$$^*OH \rightarrow {}^*O + H^+ + e^-, \quad (7)$$

$$^*O + H_2O \rightarrow {}^*OOH + H^+ + e^-, \quad (8)$$

and

$$^*OOH \rightarrow {}^* + O_2 + H^+ + e^-. \quad (9)$$

As we did for HER, we can also define the Gibbs free energy associated with each subreaction by using the adsorption energy through the equations below[5,6]:

$$\Delta G_{ads} = \Delta E_{ads} + \Delta ZPE - T\Delta S, \quad (10)$$

where ΔZPE is the change in zero-point energy, and TΔS accounts for the entropy change between the adsorbed and gaseous states at 298.15 K. Given that the zero-point energy and entropy contributions of the oxygenated intermediates are nearly identical across the different intermediates, the adsorption Gibbs free energies for the intermediates can be expressed as follows[7]:

$$\Delta G_{OH^*} = \Delta E_{OH^*} + 0.35 \text{ eV}, \quad (11)$$

$$\Delta G_{O^*} = \Delta E_{O^*} + 0.05 \text{ eV}, \quad (12)$$

and

$$\Delta G_{OOH^*} = \Delta E_{OOH^*} + 0.40 \text{ eV}. \quad (13)$$

With a $\Delta E_{ads}$ transition to each intermediate state:

$$\Delta E_{OH^*} = E_{OH^*} - E^* - [E_{H2O} - (½)E_{H2}], \quad (14)$$

$$\Delta E_{O^*} = E_{O^*} - E^* - [E_{H2O} - E_{H2}], \quad (15)$$

and

$$\Delta E_{OOH^*} = E_{OOH^*} - E^* - [2E_{H2O} - (3/2)E_{H2}]. \quad (16)$$

where E* represents the total energy of the HEA. At the same time, $E_{OH^*}$, $E_{O^*}$, and $E_{OOH^*}$ are the total energies of the catalyst with adsorbed OH, O, and OOH intermediates, respectively. $E_{H2O}$ and $E_{H2}$ denote the total energies of water and hydrogen molecules in the gas phase. We can apply a similar correction to the Gibbs free energy for the sub-reactions of OER, as we did for HER.[5]

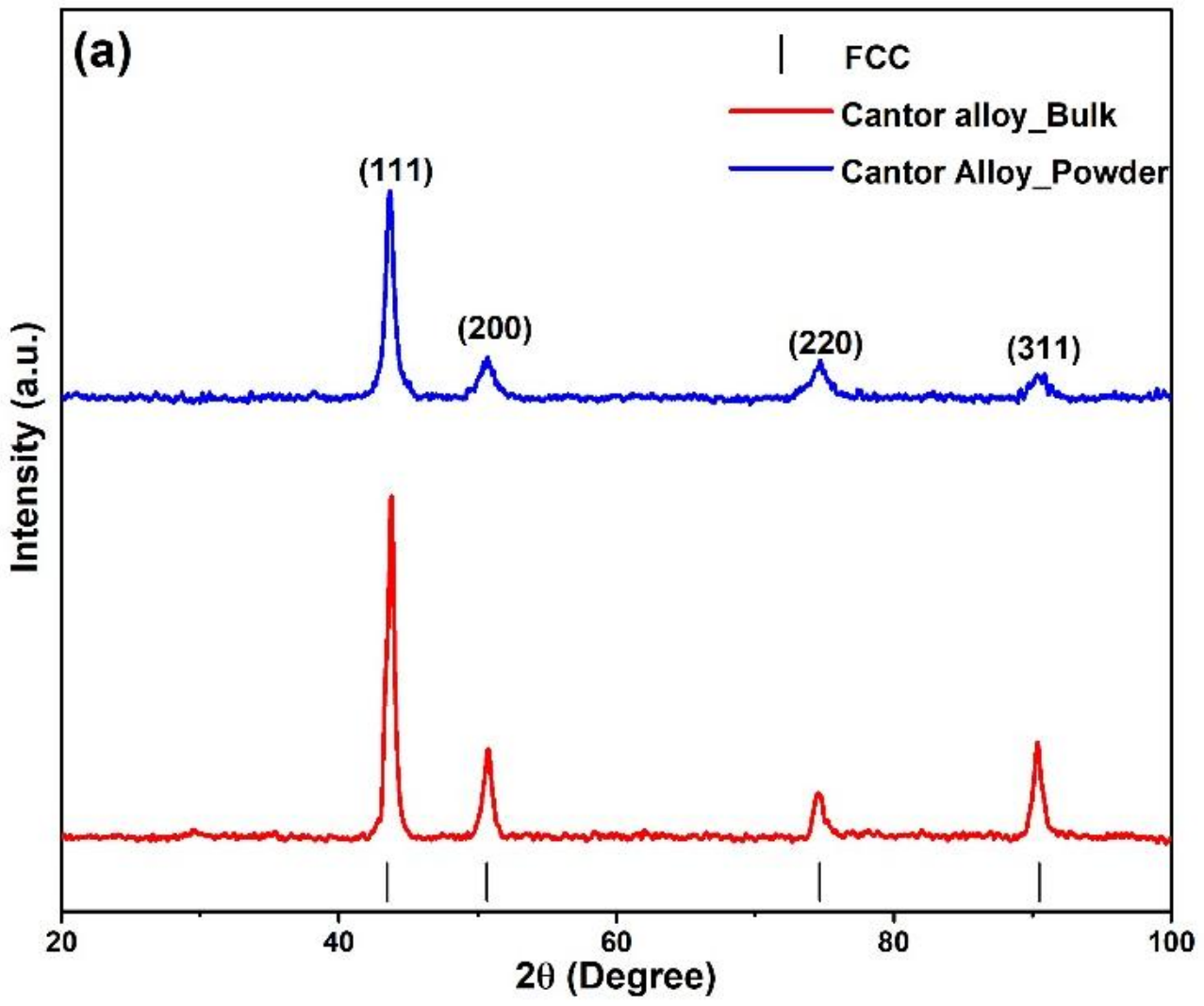

**Figure S1.** XRD of HEA-Cr high entropy alloy and Nanopowder.

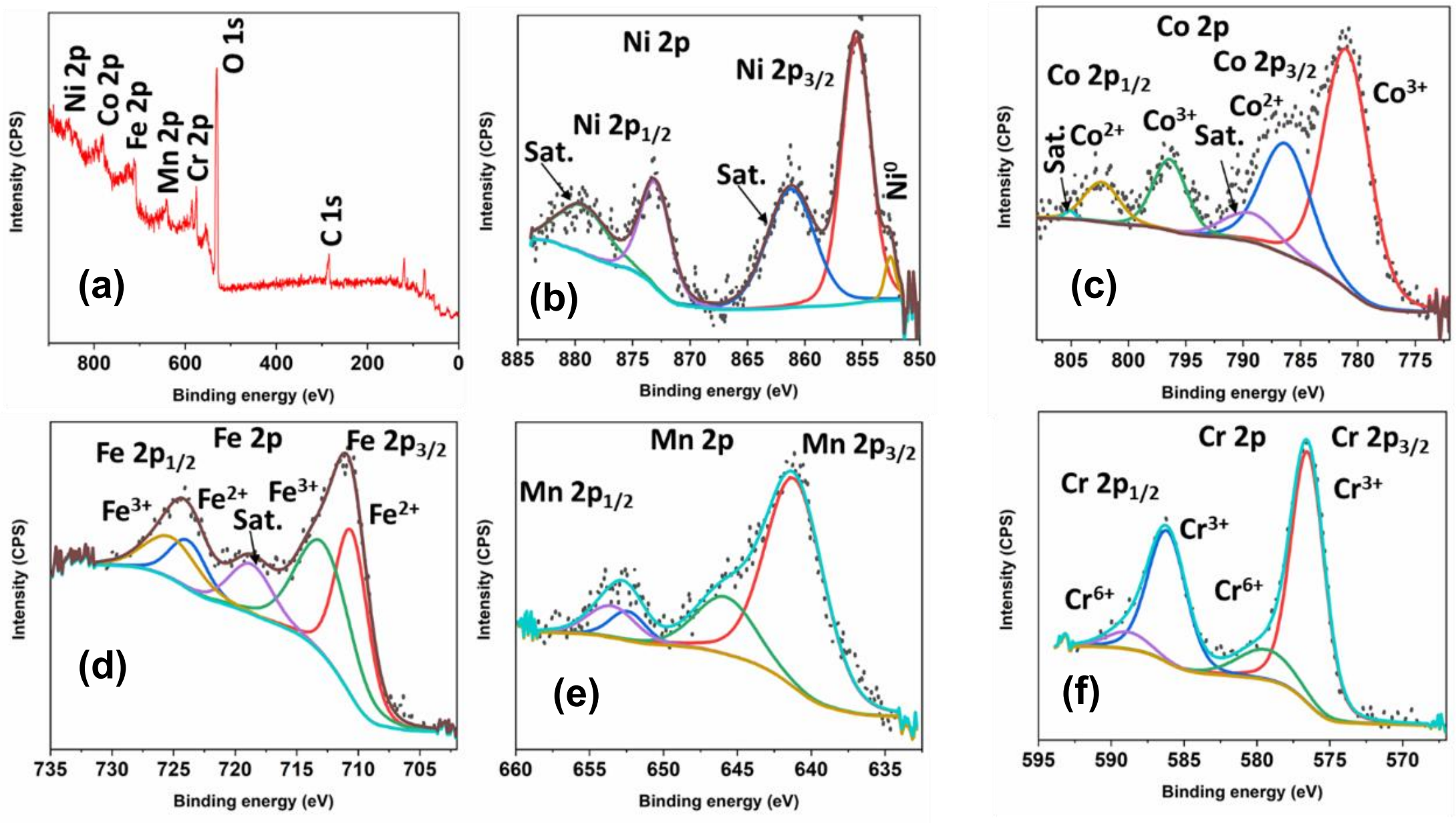

**Figure S2.** XPS analysis of HEA-Cr Cantor alloy powder showing the (a) survey scan and (b-f) deconvoluted Ni2p, Co2p, Fe2p, Mn2p, Cr2p spectra.

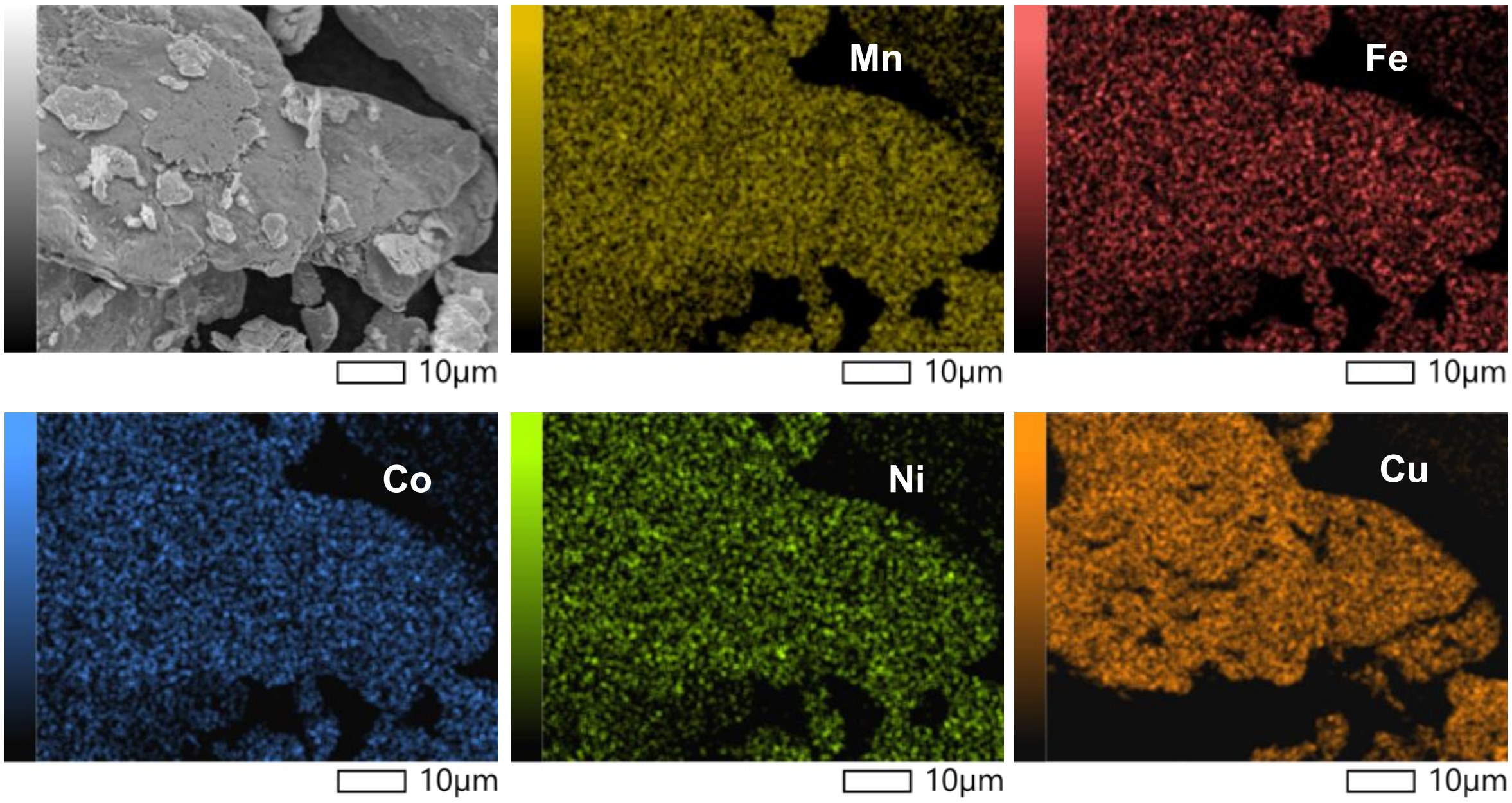

**Figure S3.** SEM EDS mapping of MnFeCoNiCu HEA catalyst.

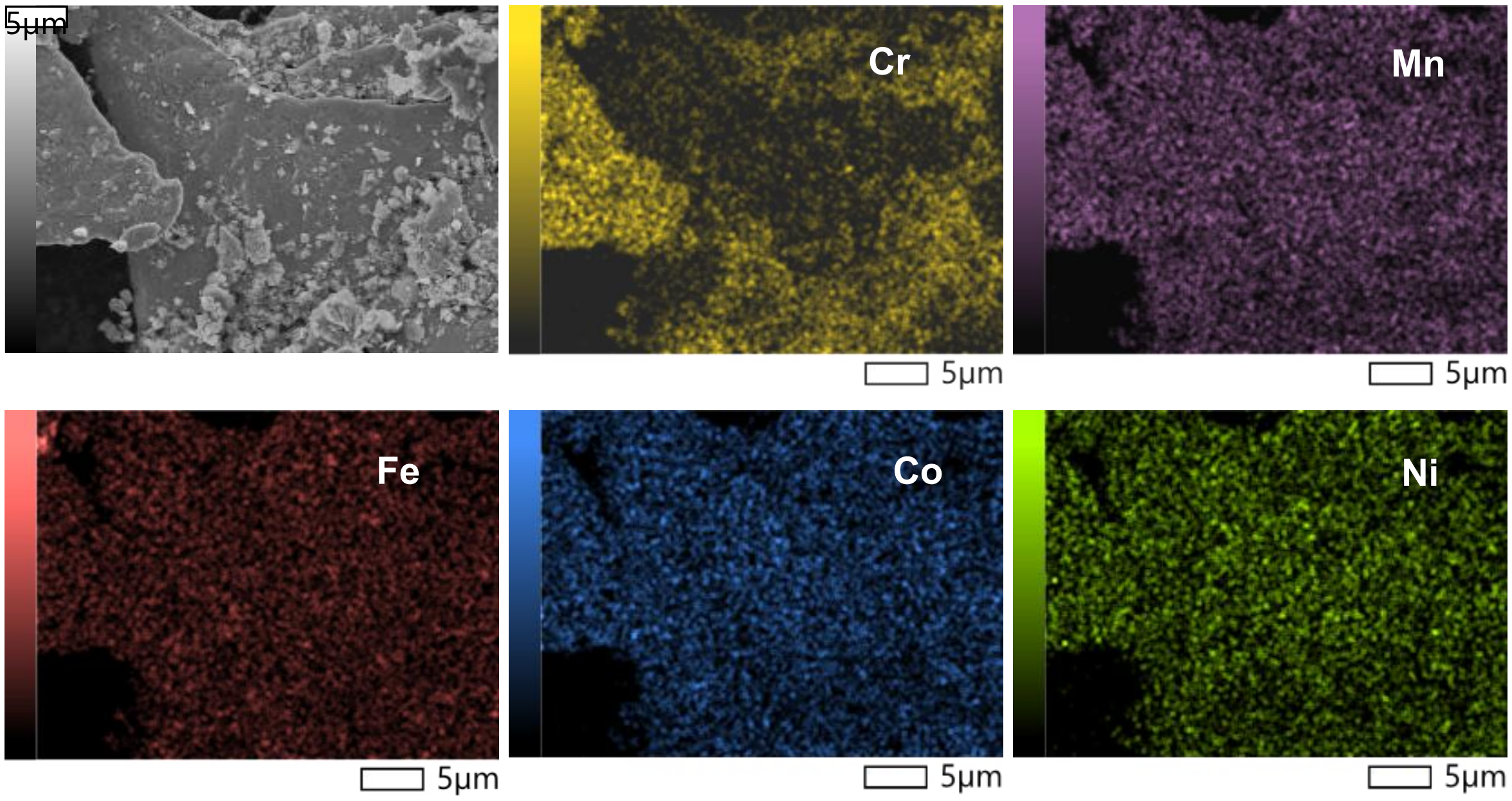

**Figure S4.** SEM EDS mapping of MnFeCoNiCr HEA catalyst.

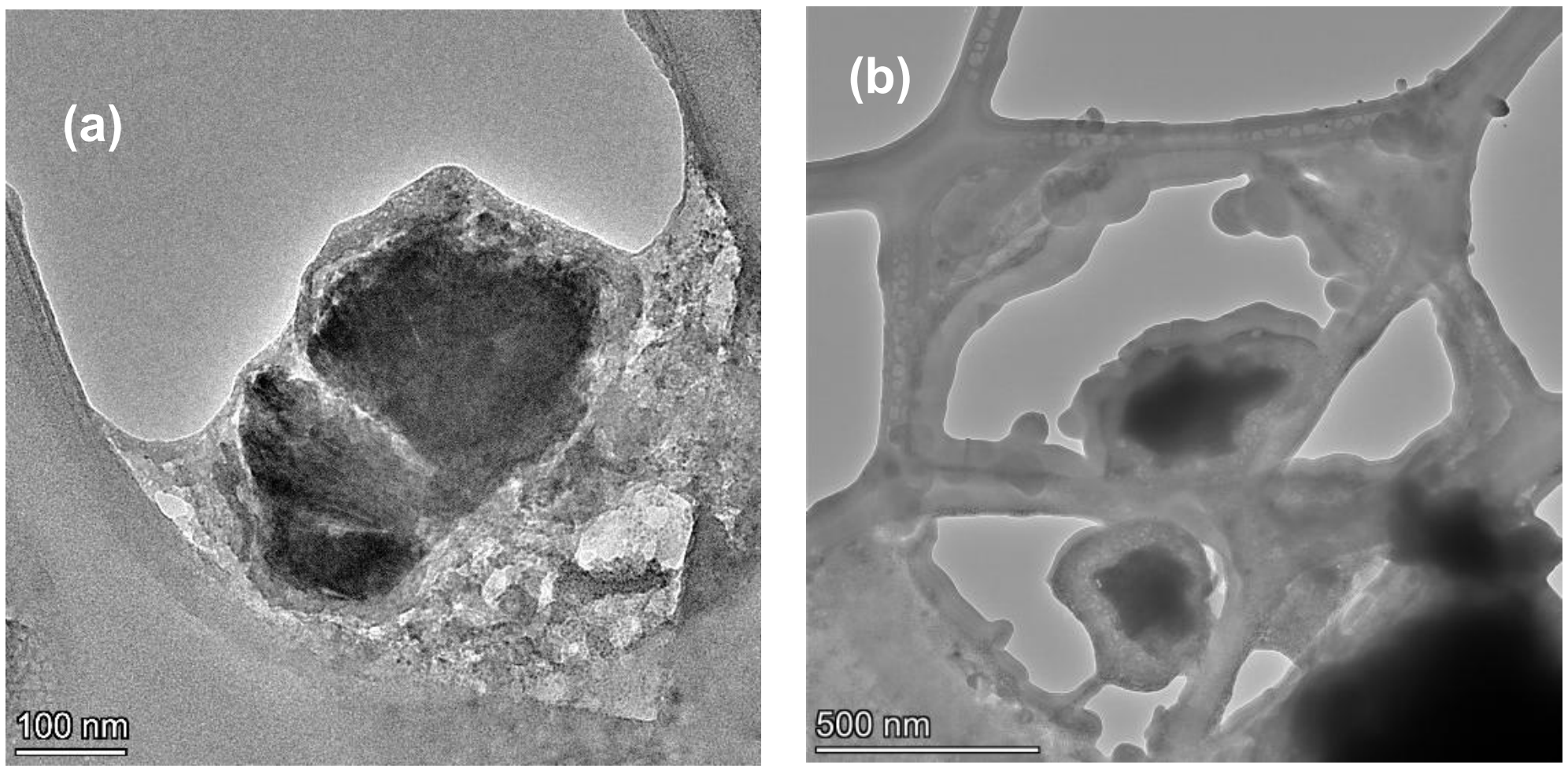

**Figure S5.** Low magnification TEM image of HEA-Cu HEA catalyst after (a) HER and (b) OER stability test (ADT). Catalyst after OER showed a present of outerlater forming a core-shell structure.

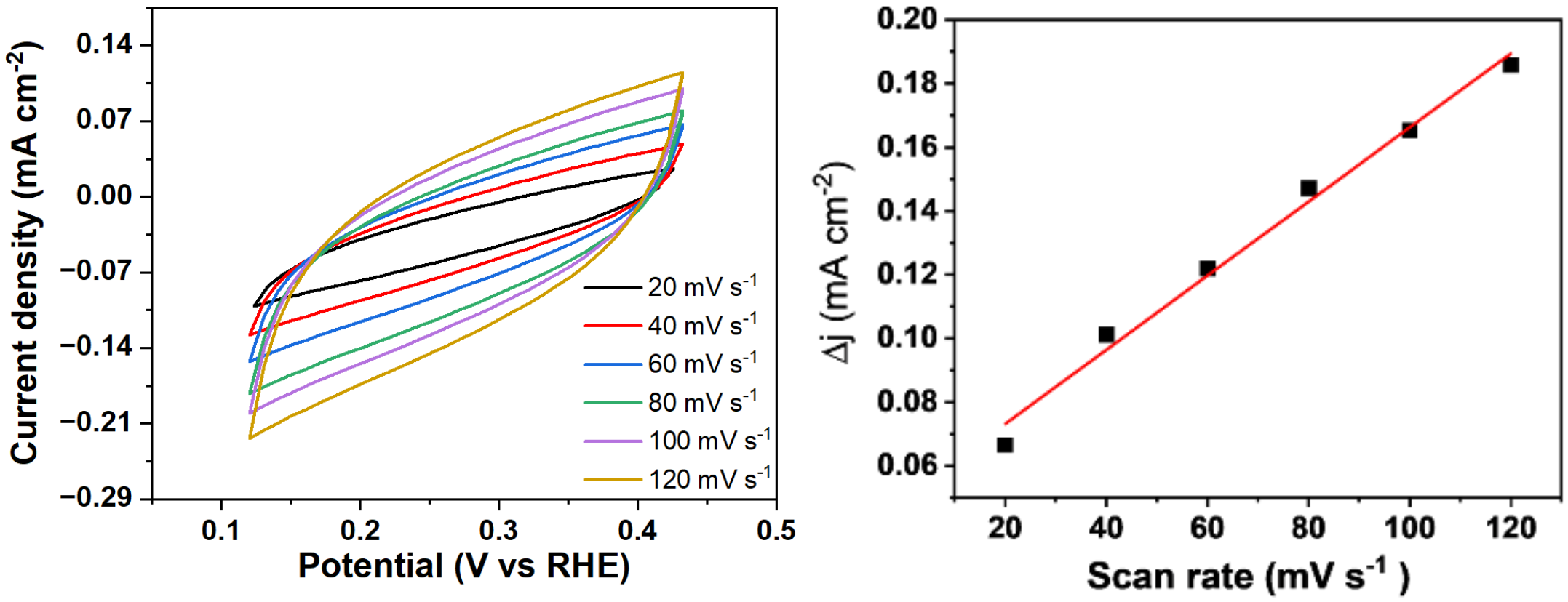

**Figure S6.** ECSA measurement: (e) CV plot at different scan rate and (f) current density vs scan rate plot.